\documentclass[ modern, longauthor ]{aastex7}     
\usepackage{xcolor}
\newcommand{\kms}{km~s$^{-1}$}
\newcommand{\etal}{et al.}

\usepackage{graphicx}

\submitjournal{Astrophysical Journal}
\received{18 November 2025 }
\revised{13 December 2025}
\accepted{16 December 2025}

\begin{document} 
\title{Betelgeuse: Detection of the Expanding Wake of the Companion Star}
\shorttitle{Betelgeuse Companion Wake}

\author[orcid=0000-0002-8985-8489,gname=Andrea K., sname=Dupree]{Andrea K. Dupree}
\affiliation{Center for Astrophysics | Harvard \& Smithsonian, 60 Garden St, Cambridge, MA 02138, United States}
\email[show]{adupree@cfa.harvard.edu}

\author[orcid=0000-0003-4019-0630,gname=Paul I., sname=Cristofari]{Paul I. Cristofari}
\altaffiliation{Center for Astrophysics | Harvard \& Smithsonian, 60 Garden St, Cambridge, MA 02138, United States}
\affiliation{Leiden Observatory, Leiden University, PO Box 9513, 2300 RA Leiden, The Netherlands}
\email[show]{cristofari@strw.leidenuniv.nl}

\author[orcid=0000-0002-1417-8024, gname=Morgan, sname=MacLeod]{Morgan MacLeod}
\affiliation{Center for Astrophysics | Harvard \& Smithsonian, 60 Garden St, Cambridge, MA 02138, United States}
\email[show]{morgan.macleod@cfa.harvard.edu}

\author[orcid=0000-0003-4484-9295, gname=Kateryna, sname=Kravchenko]{Kateryna Kravchenko}
\affiliation{Max Planck Institute for Extraterrestrial Physics, Gie{\ss}enbachstra{\ss}e 1,D-85748 Garching, Germany}
\email[show]{kkravchenko@mpe.mpg.de}
\correspondingauthor {A. K. Dupree}

\begin{abstract}
    Recent analyses conclude that Betelgeuse, a red supergiant star (HD 39801), likely has a
  companion object  with a period of $\sim$ 2000 days orbiting at only 2.3 R$_\star$, deep in the chromosphere
  of the supergiant.  A  probable  detection
  of such a companion, named  {\it Siwarha},  has just occurred from speckle imaging. This study finds that
  Betelgeuse spectra in the optical region
  and ultraviolet exhibit signatures of variable circumstellar absorption and chromospheric outflows.  These variations
  are consistent with the   $\sim$2000-day period of the companion object.  Circumstellar absorption evident in
  optical Mn I lines, and mass outflow marked by ultraviolet Fe II, Si I, and Mg I lines  increase
      after the transit of the companion across the disk of Betelgeuse. Following the  eclipse of the companion,
      the absorption and outflow slowly decrease  in advance of the next transit.  The occurrence and variation of this
      plasma appear consistent with the presence of a trailing and  expanding wake caused by a   
    companion star orbiting within the  atmosphere of Betelgeuse.

\end {abstract}

\section{Introduction}

The red supergiant star, Betelgeuse (Alpha Orionis, HD 39801) has long been known to exhibit two periods
of variation, $\sim$400 days and $\sim$2000 days, in apparent brightness,  radial velocity, astrometric measures, and
chromospheric emission lines
(R. Stothers \& K. Leung 1971; L. Goldberg 1984; E. Guinan 1984; A. Dupree \etal\ 1987;
S. Ridgway 2013).  While the 400-day period has been ascribed to
the fundamental or low-overtone  pulsational mode of the star (M. Joyce \etal\  2020),
the long secondary period (LSP)  has been attributed to a number of causes:
nonradial gravity modes, convective cells, binarity, dust formation or magnetic activity among
them (L. Kiss \etal\  2006; R. Stothers 2010; P. Wood \etal\  2004; J. Percy \&  H. Sato 2009; I. Sozynski \etal\ 2021).

Two independent studies  (J. Goldberg \etal\ 2024; M. MacLeod \etal\ 2025) conclude that the LSP 
is most likely caused by a companion star orbiting within the atmosphere of the supergiant.  J. Goldberg \etal\ (2024)
evaluated  the many suggested
explanations for the LSP and concluded that a companion star accompanied by a trailing cloud of dust could explain both the radial velocity
and photometric variations. Similar clouds of dust had been invoked by I. Soszynski \etal\ (2021) to interpret the secondary periods
observed in giant stars. M. MacLeod \etal\ (2025) assembled a century of measurements of radial velocity and
visual magnitudes as well as astrometric measures to conclude also that Betelgeuse has a companion star.   M. MacLeod \etal\ (2025)
suggest the companion has a mass of 0.6$\pm$0.14 M$_\odot$, and is orbiting at about 2.3R$_{\star}$  with its orbital plane perpendicular to the 
spin axis of the supergiant.  The focus on binarity is also attractive because it offers an explanation for the observed fast rotation rate of
Betelgeuse (H. Uitenbroek \etal\ 1998; P. Kervella \etal\  2018). 

Very recently, with speckle imaging techniques, a  probable companion to Betelgeuse has been identified (S. Howell et al. 2025). Although the
detection itself is not of high significance, the location and brightness of the companion are consistent with predictions which provide
added weight to the feature in the speckle image.  S. Howell \etal\ (2025) have named the companion {\it Siwarha} derived from the Arabic language
to align with the origin of the name Betelgeuse. This name has  just been adopted by the International Astronomical Union.\footnote  {See
https://exopla.net/star-names/modern-iau-star-names/   }

The chromosphere of Betelgeuse, as signaled by Mg II emission, extends to a diameter of at least 270 mas (H. Uitenbroek \etal\  1998) which
corresponds to about 6.4 R$_{\star}$, taking the  observed  infrared diameter at 42  mas (M. Montarg\`es  \etal\ 2014).
 Speckle imaging (S. Howell \etal\ 2025)
 in the optical region (466nm, 562nm) yields a similar diameter (41$\pm$1.2mas, 40$\pm$1.4 mas).  Thus, the companion,
orbiting  at $\sim$2.3R$_{\star}$ 
lies within the stellar chromosphere.  As M.  MacLeod et al. (2025) noted, such a configuration could lead to a  gravitationally focussed
tail or wake following the companion.  This paper seeks observational signatures of such a wake.
   
In the following sections, a brief summary of circumstellar lines in cool evolved stars (Section 2) is followed by measures of
circumstellar Mn I transitions in Betelgeuse from optical spectra (Section 3).  Ultraviolet lines indicative of mass outflow are
discussed in Section 4 and Conclusions are summarized in Section 5.

   \section{Circumstellar Lines in Supergiant Stars}
   
   The search for circumstellar features originating in the spectra of  supergiant stars began almost century ago when W. Adams and
   E. MacCormack (1935)
   noted asymmetry of several lines (Na D, Ca H and K) in the visible spectrum which were ascribed to circumstellar material.  The 
   advent of ultraviolet capability, first from balloon-borne spectrographs and subsequently the International Ultraviolet Explorer (IUE) produced
   many studies of circumstellar lines in luminous stars.  The challenge of  detecting cool circumstellar  features against
   a cool photospheric spectrum was avoided by targeting a hot companion to the primary cool star.  This companion provided a strong
   background continuum
   making it possible to detect a circumstellar absorption feature. Systems such as  $\alpha$ Scorpii (Antares), VV Cep, and AZ Cassiopeiae were
   good targets and produced many line identifications
   (K.  van der Hucht \etal\ 1980; H.-J. Hagen \etal\ 1987;  T. Kirsch \& R. Baade 1994).

   Betelgeuse presents a challenging object in which  to discern  circumstellar lines because it does not have a widely separated hot
   companion.  However, R.  Weymann (1962) in
   an extensive study of the Betelgeuse optical spectrum identified narrow circumstellar features from Fe I and Mn I  as well as five
   doubly ionized species (Ca II, Sc II, Ti II, Si II, Ba II). Circumstellar lines also  frequently appear against emission features arising
   from the chromosphere.   Prominent among these are the Mg II emission lines near 2800\AA\ which provide a variable background
   continuum to circumstellar Mn I and Fe I transitions located in the wings of Mg II.  However, details of the circumstellar
   line width and its velocity are challenging to extract  without a well-defined background emission line  profile. Thus we focus here on
   optical circumstellar transitions.
   
   Additionally the circumstellar envelope of Betelgeuse produces
   emission and this envelope has a diameter of 4 to 8 arcsec (A. Bernat \& D. Lambert 1975), and extends
   possibly as far as 50 arcsec (R. Honeycutt \etal\ 1980). Thus a spectrum obtained with an aperture
   of $\sim$ 4 arcsec or larger will contain contaminating emission.   Optical spectra  suffer from this
   contamination, and emission from the circumstellar
   shell is frequently seen close by the absorption profile.  
   

\begin{figure}
\begin{center}
\includegraphics[scale=0.49]{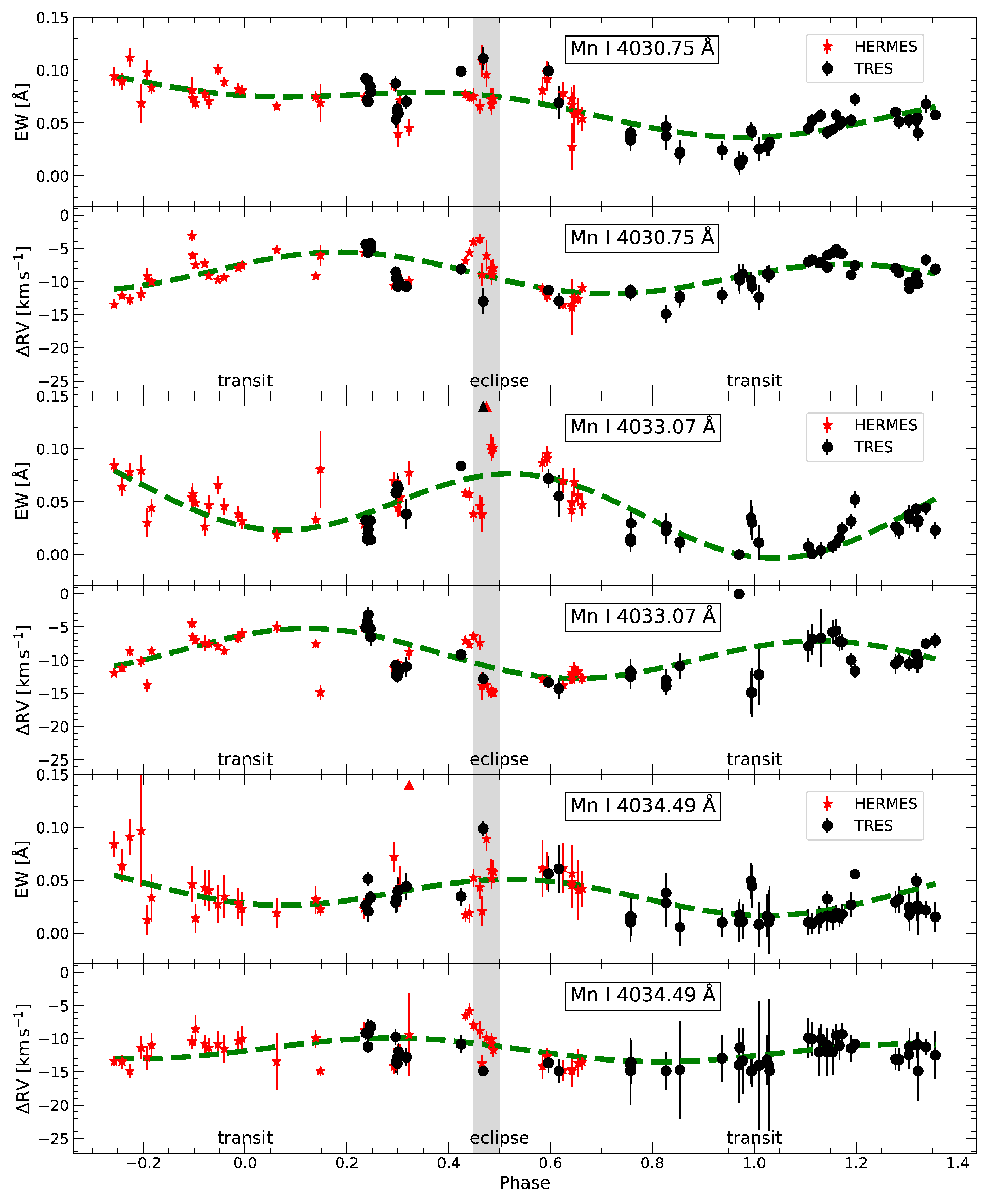}
\end{center}
\vspace*{-0.2 in}

\caption{ Measurements of three narrow circumstellar  Mn I  lines from the HERMES and TRES spectrographs.  A 2-sine curve with a period of 2109 days is
       placed to guide the eye ({\it broken green line}). Filled red or black triangles mark a few outliers. {\it Upper panel for each line:} The equivalent
       width, EW(\AA), of the Mn I lines as a function of phase of the
       companion star. Phase 0 corresponds to the transit of the companion and phase 0.5 marks the eclipse of the companion.
       The negative phasing is used here for a few of the HERMES measures because they began prior to
       the TRES measures.
       {\it Lower panel for each line:} The relative radial velocity (\kms) of the Mn I circumstellar line with respect to  the photospheric line.}
\end{figure}

   \section{Optical Circumstellar Lines in Betelgeuse}

   Optical spectra from two spectrographs are analysed here.   HERMES  is a high-resolution fibre-fed cross-dispersed echelle mounted
   on the 1.2m Mercator telescope at the Roque de Los Muchachos Observatory, La Palma (Spain) with a spectral resolution of
   86,000 and wavelength coverage from 3800\AA\ to 9000\AA\ (G. Raskin \etal\ 2011).  About 43 HERMES spectra were 
   obtained over a span of about 5.25 years:  2015 Nov 11 to 2021 May 2.  Most of these spectra were included previously in a
   detailed tomographic analysis
   of the development of shock waves in the Betelgeuse photosphere prior to the Great Dimming (K. Kravchenko \etal\ 2021).

   These spectra are  
   complemented by spectra from TRES, which is a high-throughput cross-dispersed echelle spectrograph on the 1.5m Tillinghast telescope 
   at FLWO Observatory at Mt. Hopkins, Arizona. TRES spectra cover 3900\AA\ to 9100\AA\ with a spectral resolution of 44,000.  The  55 TRES spectra 
   were obtained  starting $\sim$3 years after the start of  HERMES spectra, from 2018 Sep 18 to 2025 Mar 04 - thus extending over $\sim$6.5 years.
   The two  overlap for about 2.7 years.  Both sets of spectra offer good sampling of the 2109-day period, spanning almost two full  epochs.
   Circumstellar  transitions in Mn I also occur in the
   ultraviolet, most notably affecting the emission lines of Mg II (A. Lobel \& A. Dupree 2000).  However these ultraviolet
   transitions  appear 
   against the varying  broad centrally-reversed line profiles of the strong Mg II emission near 2800\AA, making extraction of line
   characteristics difficult. 
   
   In this analysis, the ephemeris of the companion is taken from M. MacLeod \etal\ (2025) as:
   T$_0$ (transit, phase 0) = JD 2459988.29 corresponding to 2023.12$^{+0.34}_{-0.34}$, and Period = 2109 $\pm$9  days.

  We focus on the Mn I triplet in the optical, which consists of 3 transitions (4030.75\AA, 4033.07\AA, and 4034.49\AA) arising from the
   ground level of Mn I.  To extract the line characteristics, 
   a model of the spectrum is obtained by fitting two Gaussian profiles to the circumstellar and
   photospheric line. An interactive fitting procedure is adopted to obtain an optimal fit to
   both lines. First, a single Gaussian profile is fit to the broad photospheric line, and removed
   from the observed spectrum. This allows  a first fit to the circumstellar line with
   another single Gaussian fit. To improve the fits, the modeled circumstellar line is removed from
   the original spectrum, and the process is repeated until the fitting parameters remain stable throughout iterations.
   This is shown schematically in Fig. 6 (Appendix).  Five spectra are selected to display the 3 Mn I lines and the
     Gaussian fits at 5 phases of the companion's orbit.  These are shown in Fig. 7, 8, and 9 in the Appendix.

   The equivalent width (EW) of the circumstellar line is extracted as well as the position of the line center relative to the center of
   the photospheric line.  These values are shown in Fig  1 for each of the lines where the results from HERMES are shown with
   a negative phase, because some were obtained prior to the TRES observations.

      Inspection of the changes in equivalent width of the narrow circumstellar line in the 3 Mn I transitions reveals a similar pattern
   among them.  The minimum of the equivalent width occurs near
   transit of the companion (phase 0.0 and 1.0).  The absorption continues to strengthen as the eclipse approaches (phase 0.5). After
   eclipse, the equivalent width decreases.  The HERMES spectra include observations  before the start of the  TRES  spectra and
   so represent a different epoch.  Changes in line strengths occur epoch to epoch, which is expected.  Ultraviolet and infrared
   imaging have also shown changes in emission structures across the stellar surface on time scales of months (A. Dupree \& R. Stefanik 2013;
   M. Montarg{\`e}s \etal\ 2016).

   Another measurement of interest is the velocity shift of the circumstellar line with respect
   to the  photospheric line.  This shift as determined for each transition from the Gaussian fits to the line profiles is shown in the lower panel
   of  Figure 1.  An offset of $\sim -$5 \kms\ from the photospheric velocity  increases following the transit of the
   companion, reaching maximum outflow speed around eclipse, $\sim$ $-$10 to $-$15 \kms,  and then returns to a lower value.  
   Such variations in expansion speed are characteristic of circumstellar material perhaps modulated by the presence of a 
   companion object.  Spiral structure outflows  result from radiation hydrodynamic calculations of mass-losing binary
   star models (Z. Chen \etal\ 2020).

  \section{Ultraviolet Chromospheric Lines in Betelgeuse}
   Ultraviolet line profiles are uniquely valuable in probing atmospheric dynamics.   Many transitions, such as Fe II, Si I, and 
   Mg I,  with centrally reversed emission profiles, indicate mass motions in the chromosphere by their changing asymmetries.
   These changes have been modeled in detail  to infer motion across the surface of Betelgeuse (A. Lobel \& A. Dupree 2001).

     The dynamics of the chromosphere are indicated by changes in relative strength of the line emission features. The ratio of
     the short-wavelength (``blue'')  emission peak to the long-wavelength (``red'') emission peak in a centrally reversed profile
     characterizes the motion of the atmospheric material. The ``blue'' emission peak becomes weaker relative to the ``red''
     peak when photons moving toward the observer encounter larger opacity on the blue side of the line (D. Hummer \& G. Rybicki 1968).
     Such an asymmetry indicates outflowing material.  Changes in line asymmetries and associated atmospheric velocity profiles
     have been calculated for Betelgeuse
     (A. Lobel \& A. Dupree 2001) where it  was demonstrated that increasing asymmetries in centrally
     reversed lines are caused by increasing velocities in the atmosphere.  Thus, the asymmetry and its variation provide insight
     into the velocity field.  In addition, details of the profiles, such as the  shift and broadening of  the central reversal also
     indicate mass motions at the highest regions of line formation in the atmosphere.

     We now examine the behavior of  seven ultraviolet chromospheric emission lines observed in Betelgeuse with HST/STIS  spanning the
     years 2019 to 2025.  This  time interval of  2483 days exceeds the LSP of  2109 days. 

     Transitions from the ground state or low levels of Fe II at 2382.037\AA, Mg I at 2852.1\AA, and Si I at 2516.1\AA\ are
     favorable to investigate the dynamics of the atmosphere.  These strong lines exhibit emission with a central
     reversal  that produces blue and red emission peaks.    The central reversals are broad and deep - frequently with zero flux -
     making the detection of narrow circumstellar lines difficult, even if the spectral resolution is adequate.  
     Wavelengths and  energy levels for these  ultraviolet lines are given in Table 1 and the changing  relative line strengths are
     evaluated  in the HST/STIS spectra in the  following sections.

\floattable

\begin{deluxetable}{lclccc}


\tablecaption{Selected Ultraviolet Transitions \\}
\tablehead{
\colhead{Species} &
\colhead{Wavelength (\AA)}   &
\colhead{E$_{{\rm lower}}$ (eV)} 
}

\startdata
Si I  & 2516.112 & 0.03 \\
Fe II & 2585.876 & 0.0\\
Fe II & 2692.834 & 0.99 \\
Fe II & 2724.883 & 1.04\\
Fe II & 2727.538 & 1.04\\
Fe II & 2730.734 & 1.08\\
Mg I  & 2852.127 & 0.0 \\
\enddata

\end{deluxetable}

All of the  29 HST/STIS spectra used here  were selected from the center pointings on Betelgeuse made
with the 0.025 x 0.1 arcsec aperture and the E230M grating (A. Dupree \etal\ 2020).   They sample the center of the star
which has a diameter  of 42 mas in the infrared  (M. Montarg{\`e}s \etal\ 2014) and a diameter of 125 mas in the continuum 
at 2500\AA\ (R. Gilliland \& A. Dupree 1996).   However, the Mg II emission extends farther,  exhibiting a diameter of  270 mas , 
corresponding to a radius 6.4 R$_{star}$ (H. Uitenbroek  \etal\ 1998).  The emission from a circumstellar envelope is not present 
in the  HST spectra we use, because the small spectroscopic  aperture (0.025 x 0.100 arcsec) offered by HST/STIS avoids the  
emission contributions from the circumstellar 
     envelope (A. Bernat \& D. Lambert 1975)  which has a diameter of  4 to 8 arcsec.
The  spectra  used and  line ratios are given in Appendix/Table 2,
and discussed individually in the following sections.  

\begin{figure}

  \includegraphics[scale=0.8]{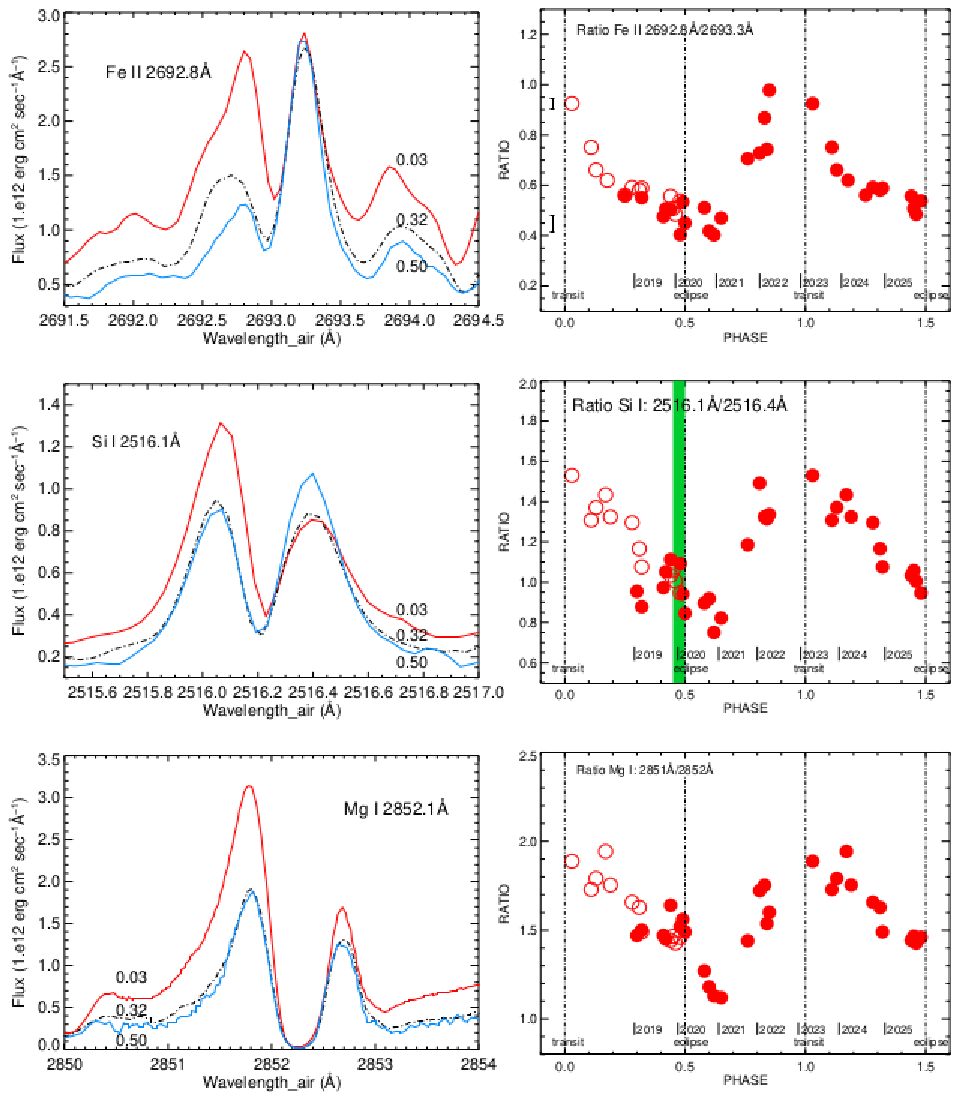}

\caption{ {\it Left panel:} Chromospheric transitions (Fe II, Si I, Mg I)  at three phases. 
       The time sequence has been inverted for display. Spectra at phases 0.03 and 0.32 were obtained in 2023 and 2024
       respectively; the spectrum at phase 0.50  was obtained in 2020.  The central absorption extending toward shorter
       wavelengths marks the expanding chromosphere after transit of the companion (phase 0).   The decrease in total
       line flux is caused by the substantial weakening of the short wavelength emission. 
       {\it Right panel:} The ratio of the peaks, (blue/red), 
       as a function of  phase of the binary.  Phases 1.0 to 1.4 (2023-2025) are repeated in the figure at
       phase 0.0 to 0.4 and marked by open circles. The green band in the middle panel marks the Great Dimming: 2019 December- 2020 February.} 

\end{figure}

\subsection{Fe II 2692.83\AA}
{\bf Three}  HST/STIS  spectra at 2692\AA\
taken at the Betelgeuse center  at different phases are displayed in Figure 2 ({\it top left panel}).
These spectra were selected from those listed in Table 2.  We take the
ratio of the maximum flux in the  short-wavelength emission feature  to that in the long wavelength emission feature (blue/red) as the parameter of 
choice to assess  the presence of asymmetries.  The decrease in total line flux from phase 0.03 to 0.50 is caused by the weakening of
the "blue" emission.

This ratio shown in Fig. 2 ({\it top right panel}) decreases post-transit, indicating that increased chromospheric outflow is present
from phase 0.0 (transit) 
to $\sim$0.7 (post-eclipse).  The outflow  reaches a maximum
value following eclipse of the companion.  Subsequent reduction of the outflow velocity returns the chromosphere to its previous state,
marking the $\sim$2000 day period. Three of the ultraviolet observations were made during the passage of
plasma from the Surface Mass Ejection in 2019 Sep.-Nov. (phase 0.41-0.44) prior to the Great Dimming in 2020 Feb. However, that ejection appeared
to occur principally in the south to south east quadrant of the star (M. Montarg{\`e}s \etal\ 2021; A. Dupree \etal\ 2020, 2022) and the
ratios of these transitions do not appear discrepant from the others.  Moreover, as this and subsequent spectra demonstrate, the pattern
is similar during 2025 when a Great Dimming event has not occurred.

The shape of the central reversal can also indicate chromospheric motions at the highest levels of line formation.  In Fig 2 ({\it left panels}),
the reversal has a shift of the short wavelength wing post-transit in agreement with the inference from the line emission ratios. The greater
short wavelength extent  ($\sim$0.18\AA) of the central absorption at phase $\sim$0.5, an indicator of mass outflow,
corresponds to $\sim$ $-$20 \kms.
This value is somewhat larger than the values indicated by  the distant circumstellar  features,  but not surprising as the Betelgeuse wind does not
appear to be similar to the solar wind that accelerates to supersonic velocities (S. Cranmer \& A. Winebarger 2019).

\subsection{Si I 2516.1\AA}

This transition arises from a low lying level (0.03 eV) of Si I,  and the mass motions also  affect the
short-wavelength emission of the transition (Fig. 2, middle panel).  This transition was evaluated in detail at various positions on the Betelgeuse
disk which demonstrated the impact of a changing velocity field on the line asymmetries (A Lobel \& A. Dupree 2001).  We construct the
ratio of the flux of the blue emission peak to that of the red emission peak: 2516.1\AA/2516.4\AA.  This blue/red ratio shows a 
variation similar to that of the Fe II line discussed previously.  Moreover as shown in the left panel of Fig. 2, the centrally
reversed profile is broadened towards  shorter wavelengths by $\sim$0.06\AA,  in the spectra taken at phase 0.32 and 0.50.  This shift amounts
to $\sim$ $-$7 \kms.  

cd 
\subsection{Mg I  2852.1 \AA}

The transition from the ground state of Mg I is  also a likely candidate to track the circumstellar material. However the broad central reversal
spanning $\sim$0.2\AA\ or $\sim$21~\kms\  with essentially zero flux, prevents a clear detection of a circumstellar transition.   A  slight weakening appears
on the short wavelength side of the  centrally reversed chromospheric profile (Fig. 2, lower panel).
The ratio of the short wavelength:long wavelength peak  tracks the changing absorption in the outflowing chromosphere similar to that in Si I
and Fe II.

\begin{figure}[ht]!
  
\begin{center}
  \includegraphics[angle=0, scale=0.8]{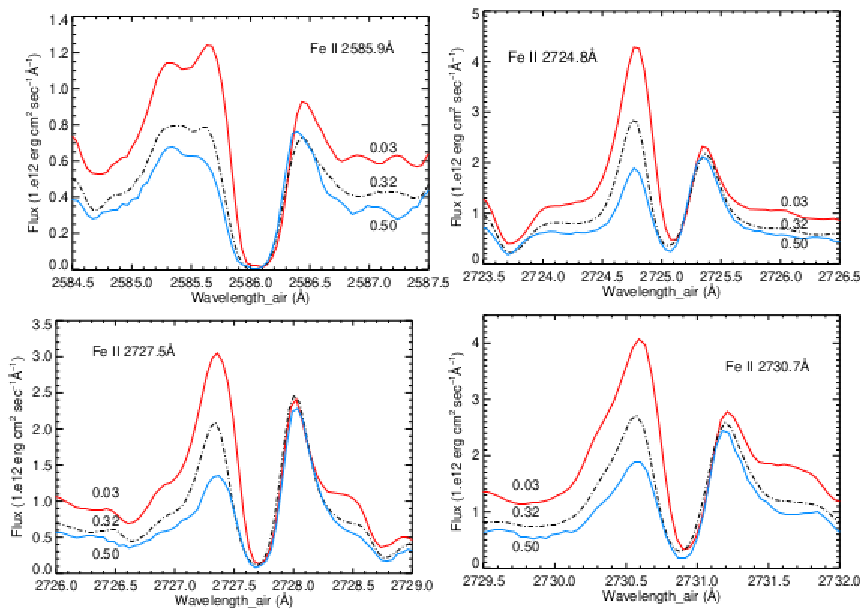}
  \end{center}

\caption{Four  Fe II transitions at three phases of the LSP. 
       The time sequence has been inverted for display as in Fig. 2.   Absorption extending toward shorter wavelengths indicates
       the expanding chromosphere after transit of the companion (phase 0).  The asymmetry of the profiles changes with phase,
       displaying  the weakening in  the blue emission  produced by the opacity in the  expanding atmosphere  and signaled by a 
       decrease in the blue/red ratios.}

\end{figure}

\begin{figure}[hb]
  \begin{center}
  \includegraphics[angle=90, scale= 0.48]{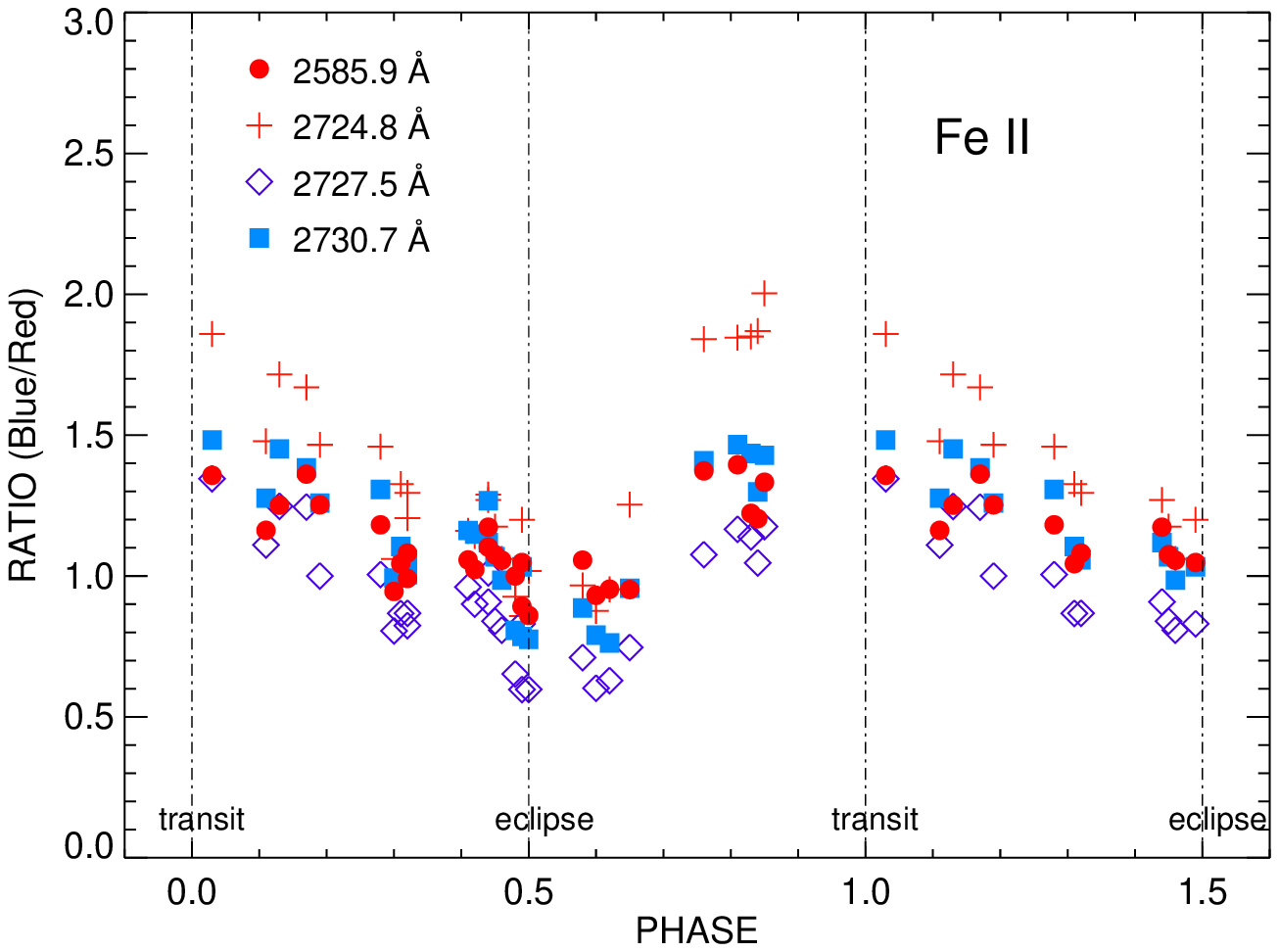}
  \end {center}

  \vspace*{-0.2 in}
  
  \caption{The asymmetry ratio (blue/red) for four Fe II transitions whose profiles are shown in Fig. 3.}
\end{figure}

\subsection{Additional Fe II lines}
To ensure that the 3 transitions of  Si I, Mg I and Fe II were not anomalous, we evaluate 4 additional lines ranging in
excitation potential from 0 to 1.08 eV. Line profiles at 3 phases are shown in Fig. 3 where a short wavelength expansion
of the central reversal post-transit
is visible and amounts to $\sim$ $-$9 to $-$12 \kms. The blue/red ratios for these transitions are shown in Figure 4.  The pattern of increased
outflow post-transit is found in all lines and replicates those discussed previously.

\section{Conclusions}

The EW changes in the optical  circumstellar Mn I lines demonstrate variability consistent with the LSP.   Adopting the model that a
companion object causes the LSP variation, we find that the phasing indicates that  circumstellar absorption   increases 
immediately following  the transit of the companion across the disk of Betelgeuse (phase 0.0), reaching a maximum at 
about phase 0.5 and then begins to decrease, returning to its initial state after $\sim$2100 days.  This pattern is similar
to the optical (V) variation found in the LSP of Betelgeuse, namely brightest at phase 0 and faintest at phase 0.5 (M. MacLeod \etal\  2025).
Thus it  appears
possible that the variation of the circumstellar material contributes to the optical variation.

In addition, the  circumstellar lines  display a variation in expansion velocity relative to the photosphere, consistent with the LSP.
As Fig. 1 shows,   following transit (phase 0.0) the circumstellar outflow
from the photosphere begins, reaching a maximum value of $\sim -$5 to $-$10 \kms\ near phase $\sim$0.7.

The  asymmetry changes
in  the ultraviolet emission line profiles signal that outflow of the chromosphere also begins post-transit
and continues through eclipse.  The chromospheric expansion is present until
phase $\sim$0.7 when it decreases rapidly as transit approaches, signaling a weakening of the outflowing motion. 
This velocity variation  has a period approximately
equal to the LSP, and it is asymmetric, reaching a maximum outflow at phase $\sim$0.7  followed by a rapid recovery as
transit approaches.  The broadened shift of the
central reversal of the chromospheric lines  ranges from $-$6 to $-$20 \kms , values commensurate with the circumstellar 
features. Thus the chromosphere and the circumstellar material appear to move in similar fashion over the LSP.

If the postulated and probably-detected companion to Betelgeuse is confirmed (J. Goldberg \etal\ 2024; M. MacLeod \etal\ 2025;
S. Howell \etal\  2025), 
we might understand the spectroscopic variations described here as follows.  A companion orbiting in a $\sim6$ year period around Betelgeuse moves 
at an orbital velocity of $\sim 43$~\kms\ (M. MacLeod \etal\ 2025). The chromosphere region of Betelgeuse is a mixed environment of multi-phase gas 
ranging from several to ten thousand Kelvin in temperature. At a representative $\sim2500$K temperature for the intermediate temperature gas 
(E. O’Gorman \etal\  2020), the sound speed is $\sim6$ \kms\  and the motion of the companion has a relative Mach number through the stellar wind 
of $\cal M \sim$ 7.  In the chromosphere, the expansion velocity of the wind is similar to that of the circumstellar absorption, on the
order of 10 \kms\  while the turbulent velocity is even higher, 
$\sim 20$ \kms\ (A. Lobel  \& A.  Dupree 2001; N. Soker 2021; S. Fuller \& D. Tsuna 2024;  J-Z. Ma \etal\  2025). The companion would
gravitationally focus surrounding 
wind into its vicinity, forming a trailing wake, whose primary orientation would be along the direction of the orbital
path (Z. Chen \etal\ 2020, see Figure 5).  
Within this wake, denser, shocked gas would be swept up and accumulate behind the passing shockwave (E. Ostriker 1999).
We should expect this wake to be quite variable because of the multi-temperature turbulent gas it exists in, and because the
turbulent velocity is larger than the mean sound-speed and a large fraction of the orbital speed.

\begin{figure}
 \begin{center}
  \includegraphics[angle=0, scale= 0.4]{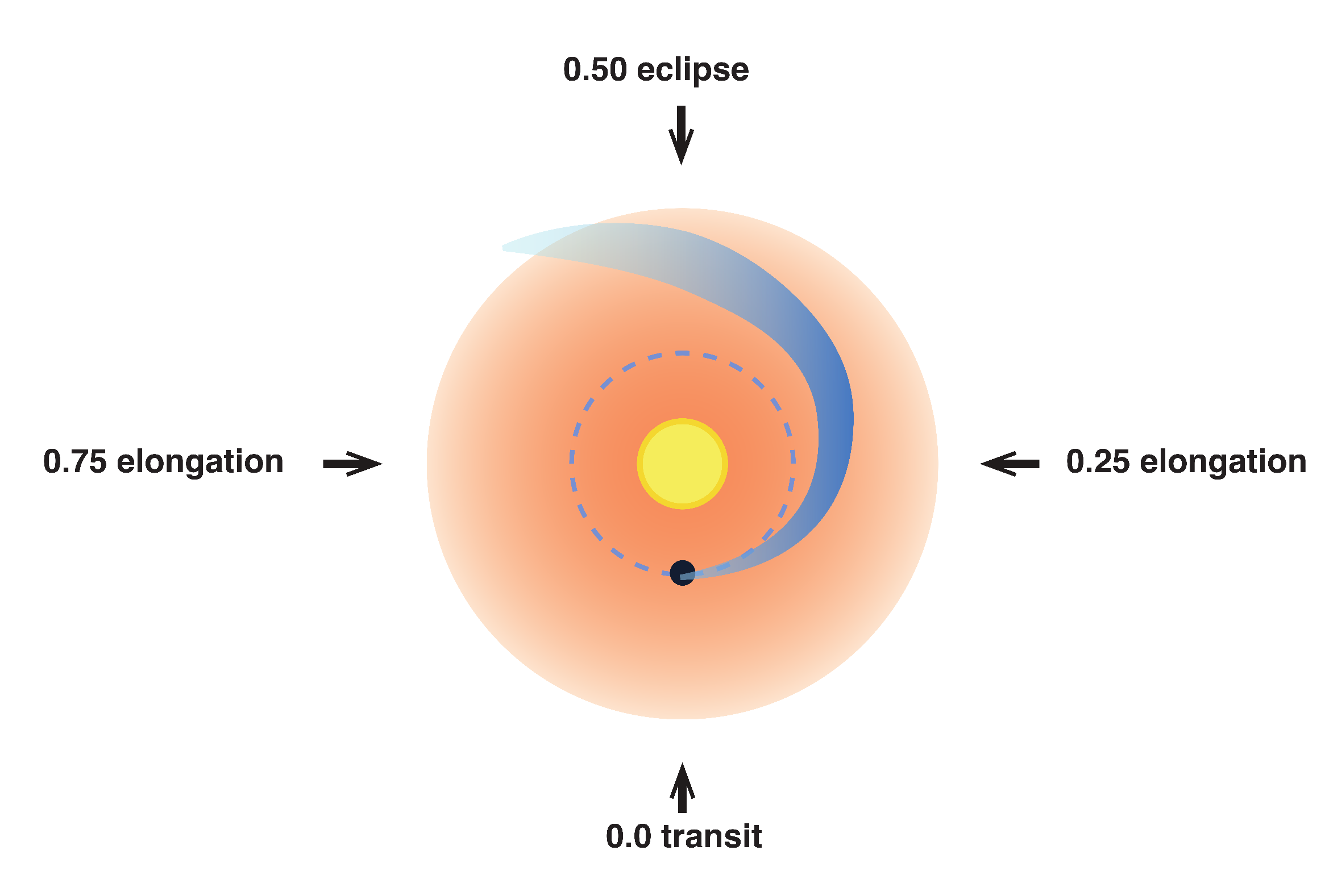}
  \end {center}
 \caption { A schematic drawing from above of Betelgeuse and the observed wake caused by the companion star.  The
     orbit of the companion, at $\sim$2.3 R$_\star$ is
   shown by a broken blue line. It is well within the total extent of the Mg II emission which reaches 6.4R$_\star$ - marked by diffuse orange color.
   When observed at different phases, along the direction
   of the arrows, the expanding wake can be detected.} ({\it Credit: A.H. Szentgyorgyi})

\end{figure}

Lines of sight passing through the wake would pass through denser, shocked material. The wake expands laterally at the sound speed, 
$\sim 6$ \kms. Thus while it is small during transit of the companion across Betelgeuse’s disk, it widens and covers a larger fraction
of the disk as 
the orbit progresses. The estimated time to cover Betelgeuse’s radius is $\sim 800 R_\odot / 6$ \kms  $\approx 3$~yr.  During this time, the 
companion has completed a half-orbit from transit to the vicinity of eclipse. Compared  to our observational data, a half-orbit offset from transit 
(i.e. eclipse at phase 0.5) is the time of minimum optical brightness in the LSP and comparable to the  most significant excess
chromospheric absorption of both optical and UV lines in these current observations. As the orbit progresses further, the wake still
exists, but as it widens, its effect is diluted by  its continued expansion (E. Ostriker 1999). Clearly, more sophisticated modeling,
perhaps capturing the thermodynamic properties of the chromosphere region
 (e.g. B. Freytag \&  S. H\"ofner 2008; S. Wedemeyer \etal\ 2017; J-Z. Ma \etal\ 2025) along with the effect of an orbiting companion 
(e.g. Z. Chen \etal\  2020), is eventually needed. In particular, the shock heating and radiative cooling and the confluence of the
gravitationally-focused wake will determine the wake density contrast, along with its characteristic temperature. These factors are
crucial to predicting the emergent spectra and are best modeled in a hydrodynamic context.

These results add  direct spectroscopic  evidence  that the extended atmosphere of Betelgeuse is modulated on timescales associated with the
LSP.  This evidence supports the  hypothesis of the presence of a companion object in the atmosphere of Betelgeuse.  In particular, 
the fact that the photosphere, the chromosphere, and circumstellar lines all appear to vary in intensity and radial velocity in synchronization with 
this long period does appear to be best explained by a companion star orbiting within the extended atmosphere.  We suggest that a trailing
wake geometry is most consistent with the observations, and calculations of the wake structure and the emergence of spectral lines from it
are needed.   Because the companion is separated by only $\sim$2.3 R$_{\star}$, it is obscured 
by the large Betelgeuse disk at the current phase $\sim$0.5, and will be visible again at phase $\sim$ 0.6 which occurs in August 2027. 

\section {Data Availability}
  Most of the   HERMES spectra are published in K. Kravchenko \etal (2021) and  are  available at the CDS via anonymous ftp
  to cdsarc.u-strasbg.fr (ftp://130.79.128.5) or via http://cdsarc.u-strasbg.fr/viz-bin/cat/J/A+A/650/L17.
The HST data described here may be obtained from the MAST archive at \dataset[doi: 10.17909/zz3e-gh10]  .
Remaining spectra from HERMES and those of  TRES are available from the authors. 

\facilities{HST(STIS), 1.2m Mercator telescope (HERMES), FLWO(TRES)} 

\software{astropy, IDL, IRAF, Adobe Photoshop}

\begin{acknowledgements}
  We are grateful to the referee whose suggestions improved the presentations found here.
  Based on observations made with the Mercator Telescope, operated on the island of La Palma by the Flemish Community, at the Spanish
  Observatorio del Roque de los Muchachos of the Instituto de Astrofísica de Canarias. These  spectra were obtained with the HERMES
  spectrograph, which is supported by the Research Foundation - Flanders (FWO), Belgium, the Research Council of KU Leuven, Belgium, the
  Fonds National de la Recherche Scientifique (F.R.S.-FNRS), Belgium, the Royal Observatory of Belgium, the Observatoire de Genève,
  Switzerland and the Thüringer Landessternwarte Tautenburg, Germany.  This research is also based on ultraviolet observations made
  with the NASA/ESA Hubble Space Telescope obtained from the Space Telescope Science Institute, which is operated by the Association
  of Universities for Research in Astronomy, Inc., under NASA contract NAS 5–26555. These observations are associated with programs
  GO 15641, 15873, 16216, 16655, 16984, 17522, and 17845.  This work was supported in part by STScI Grant HST-GO-15641.001-A to the Smithsonian
  Astrohysical Obseratory.
\end{acknowledgements}
\begin{contribution}
  HST/STIS  and TRES specta were obtained by A. Dupree; K. Kravchenko obtained the HERMES spectra; P. Cristofari and A. Dupree
analyzed the spectra. M. MacLeod contributed to the theoretical interpretation. All of the authors contributed to the discussion and conclusion.
\end{contribution}

\appendix

\begin{figure}[ht]

  \begin{center}
  
  \includegraphics[scale=0.45]{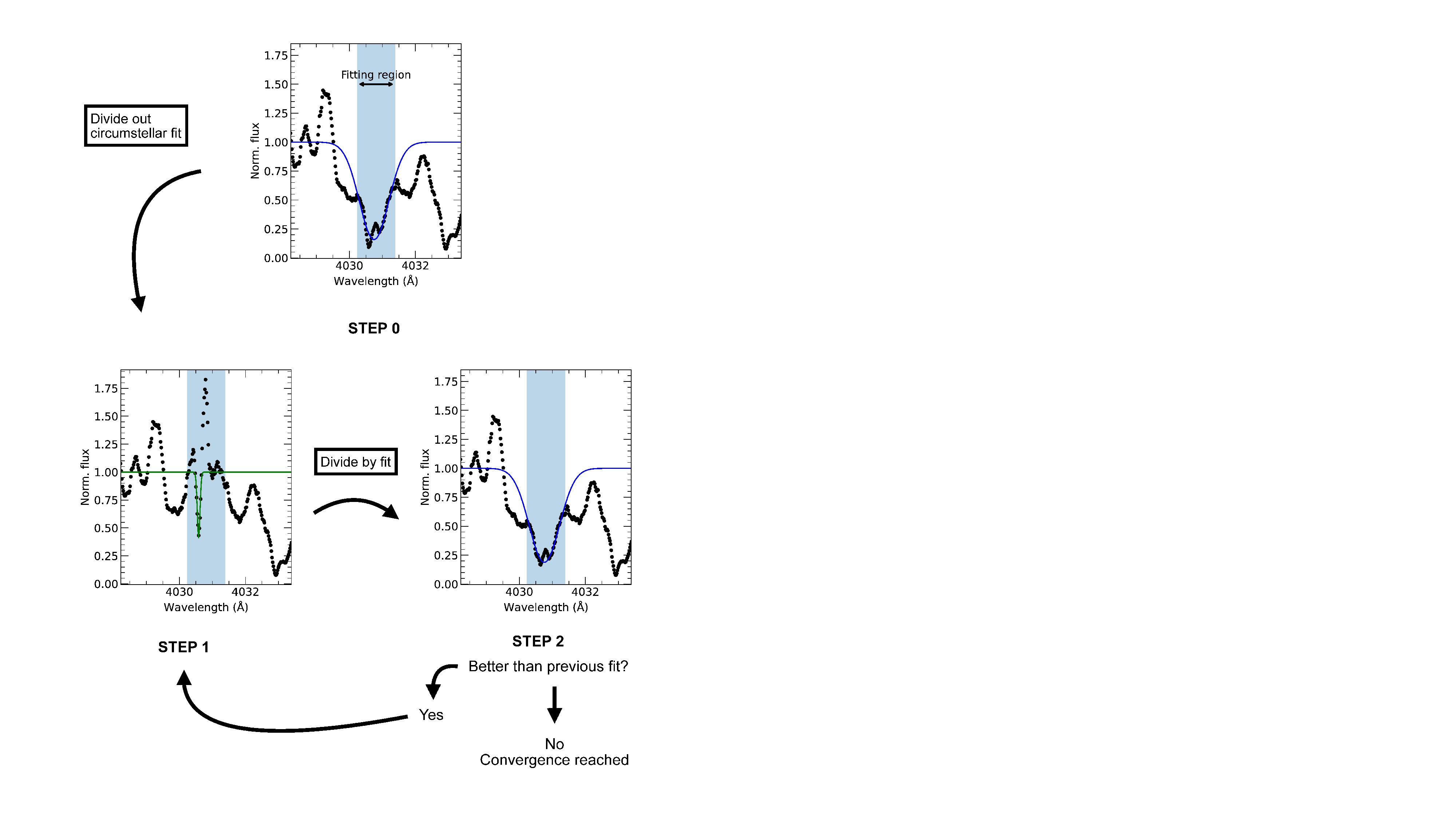}
  \end{center}
  \vspace*{-0.5 in}
  
  \caption{A HERMES spectrum of the Mn I transition at  4030.75\AA\ illustrating the
    process of extracting the narrow circumstellar absorption profile. Step 0: a Gaussian profile ({\it blue line}) is
    fit to the broad photospheric absorption line. Step 1: A Gaussian profile is fit to the circumstellar profile ({\it green line})
    after   dividing by the photospheric Gaussian fit in Step 0.  Step 2: The original line profile is divided by the
    fit to the circumstellar line (Step 1) to obtain the photospheric line profile, and a new Gaussian fit to the photospheric
    line is obtained ({\it blue line}).   If this iteration improves the $\chi^2$ between the
    model and the observed line, the process is repeated from Step 1 until no change occurs in the profile.  Both the circumstellar line
    center and its EW are estimated from the Gaussian fit to the circumstellar line.  The photospheric line center is taken from
  the Gaussian obtained in the final iteration.}
   
    \end{figure}

\begin{figure}[ht!]
  \begin{center}
     \includegraphics[scale=0.6]{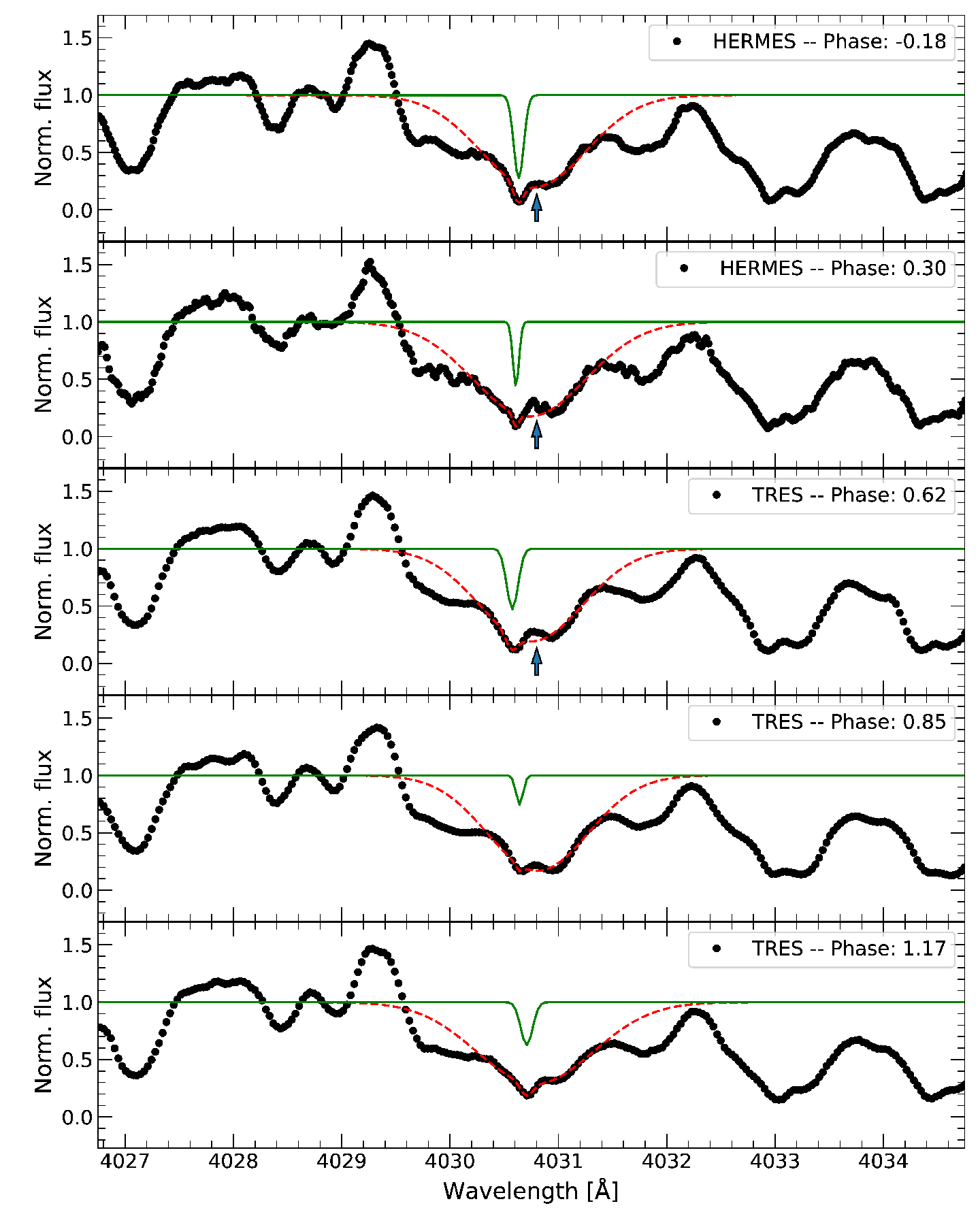}
\end{center}
  \caption{Sample spectra of the Mn I transition at 4030.75\AA\ from the HERMES and TRES observations. 
    The narrow   green line denotes the circumstellar absorption feature.
    The curve marked in red displays the fit to both features.
     Circumstellar emission can be seen in the core of the photospheric line and is marked by the upward arrow.}
   \end{figure}

\begin{figure}[ht!]
  \begin{center}
\includegraphics[scale=0.7]{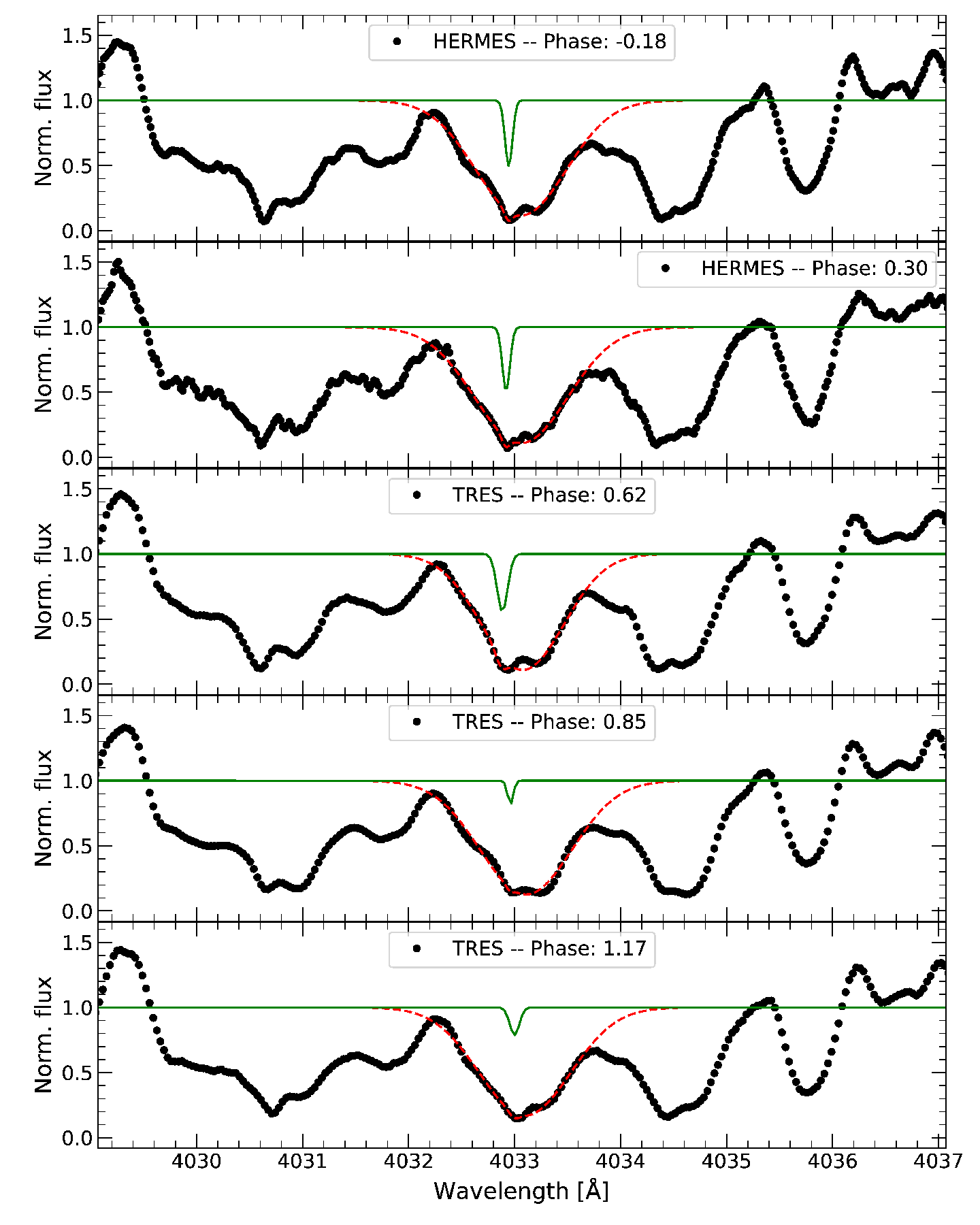}
\end{center}
\caption{Spectra of the Mn I transition at 4033.07\AA\ from HERMES and TRES observations. The curve marked in red
       displays the  fit to the blended feature,  and the narrow green line denotes the circumstellar absorption feature.}
\end{figure}
\begin{figure}[ht!]
  \begin{center}
\includegraphics[scale=0.7]{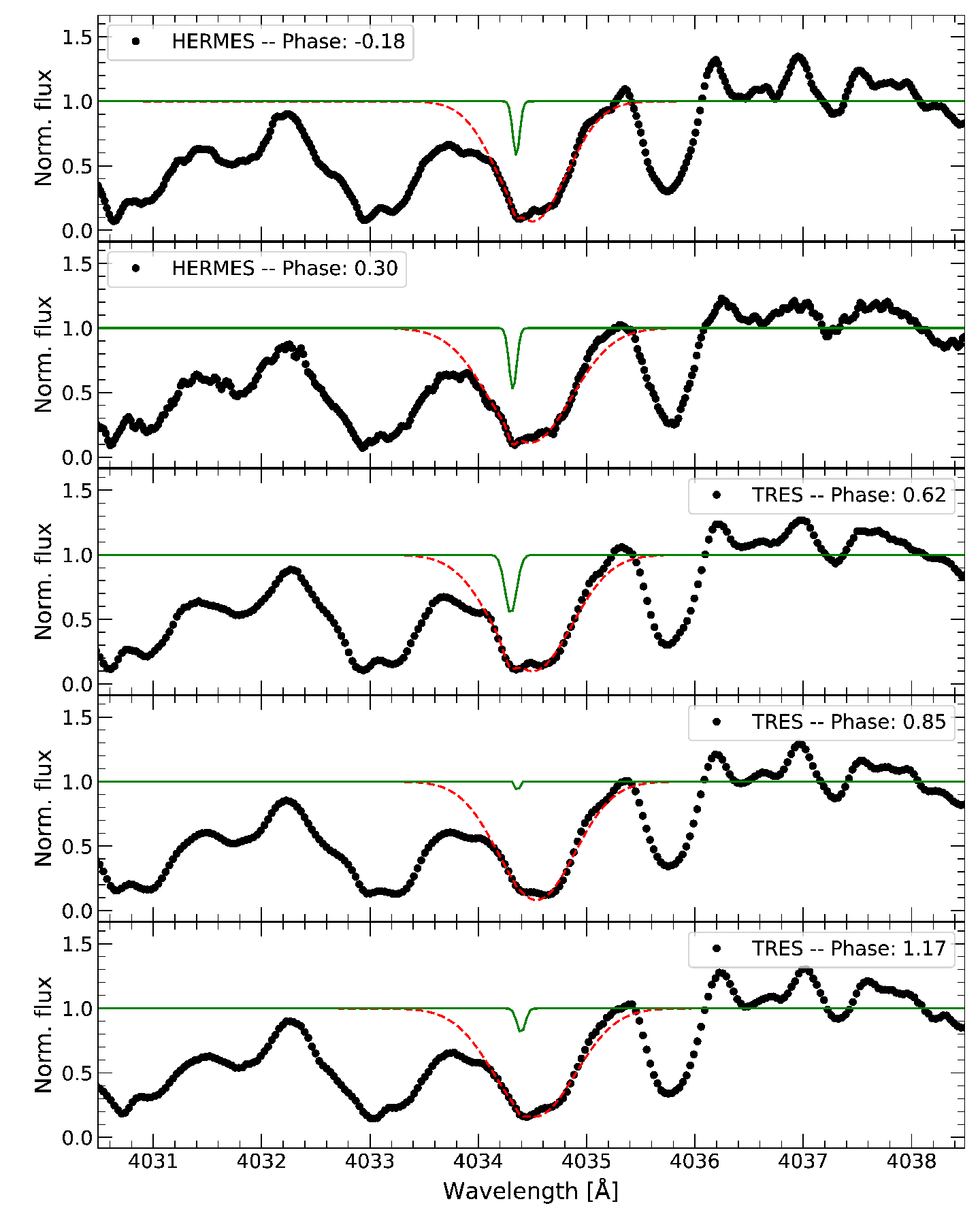}
\end{center}
\caption{Spectra of the Mn I transition at 4034.49\AA\ from HERMES and TRES observations. The curve marked in red
       displays the  fit to the blended feature,  and the narrow green line denotes the circumstellar absorption feature.}
\end{figure}

\newpage

\floattable

\begin{deluxetable}{lclcccccccccc}
\tablecaption{STIS Spectra and Line Ratios}
\tablehead{
\colhead{Date} &
\colhead{JD}   & 
\colhead{Phase} & 
\colhead{STIS spectrum} &
\colhead{Ratio Fe II}&
\colhead{Ratio Mg I}&
\colhead{Ratio Si I}&
\colhead{Ratio Fe II}&
\colhead{Ratio Fe II}&
\colhead{Ratio Fe II}&
\colhead{Ratio Fe II}\\
\colhead{ }&
\colhead{ }&
\colhead{ }&
\colhead{ }&
\colhead{2692.8\AA}&
\colhead{2852.1\AA}&
\colhead{2516.1\AA}&
\colhead{2585.9\AA}&
\colhead{2724.8\AA}&
\colhead{2727.5\AA}&
\colhead{2730.7\AA}&
}
\startdata
2019-01-26 &2458509.5  &  0.30 & odxg01040  & 0.56  &  1.47 & 0.95 & 0.95&1.06 &0.81 &0.99 \\
2019-03-05 &2458545.5  &  0.32 & odxg02040  & 0.55  &  1.50 & 0.88 & 0.99&1.21 &0.82 &1.01 \\
2019-09-18 &2458744.5  &  0.41 & odxg07040  & 0.48  &  1.47 & 0.97 & 1.06&1.16 &0.96 &1.16 \\
2019-10-06 &2458762.5  &  0.42 & odxg08040  & 0.51  &  1.45 & 1.05 & 1.02&1.14 &0.90 &1.15 \\
2019-11-28 &2458815.5  &  0.44 & oe1i01040  & 0.50  &  1.64 & 1.11 & 1.10&1.29 &1.01 &1.27 \\
2020-02-03 &2458882.5  &  0.48 & oe1i52040  & 0.40  &  1.52 & 1.09 & 1.00&0.93 &0.65 &0.81 \\ 
2020-02-25 &2458904.5  &  0.49 & oe1i03040  & 0.53  &  1.56 & 0.94 & 0.89&0.86 &0.60 &0.79 \\
2020-04-02 &2458941.5  &  0.50 & oe1i04040  & 0.45  &  1.49 & 0.84 & 0.86&1.02 &0.60 &0.78 \\
2020-08-31 &2459092.5  &  0.58 & oedq01040  & 0.51  &  1.27 & 0.90 & 1.06&0.97 &0.71 &0.89 \\
2020-10-15 &2459137.5  &  0.60 & oedq02040  & 0.42  &  1.18 & 0.92 & 0.93&0.88 &0.60 &0.79 \\
2020-11-24 &2459177.5  &  0.62 & oedq52040  & 0.40  &  1.13 & 0.75 & 0.95&0.95 &0.63 &0.76 \\
2021-02-09 &2459254.5  &  0.65 & oedq03040  & 0.47  &  1.12 & 0.82 & 0.95&1.25 &0.75 &0.96 \\
2021-09-18 &2459475.5  &  0.76 & oedq54040  & 0.71  &  1.44 & 1.19 & 1.37&1.84 &1.08 &1.41 \\
2022-01-10 &2459589.5  &  0.81 & oen702030  & 0.73  &  1.72 & 1.49 & 1.39&1.85 &1.17 &1.47 \\
2022-02-10 &2459620.5  &  0.83 & oen703030  & 0.87  &  1.75 & 1.32 & 1.22&1.85 &1.14 &1.44 \\
2022-03-11 &2459649.5  &  0.84 & oen704030  & 0.74  &  1.54 & 1.31 & 1.20&1.87 &1.05 &1.30 \\
2022-04-11 &2459680.5  &  0.85 & oen705030  & 0.85  &  1.60 & 1.33 & 1.33&2.00 &1.18 &1.43 \\
2023-04-18 &2460052.5  &  1.03 & oen751030  & 0.92  &  1.89 & 1.53 & 1.36&1.86 &1.35 &1.48 \\
2023-09-27 &2460214.5  &  1.11 & oevc01030  & 0.75  &  1.73 & 1.31 & 1.16&1.48 &1.11 &1.28 \\
2023-11-13 &2460261.5  &  1.13 & oevc02030  & 0.66  &  1.79 & 1.37 & 1.25&1.72 &1.25 &1.45 \\
2024-02-13 &2460353.5  &  1.17 & oevc03030  & 0.62  &  1.94 & 1.43 & 1.36&1.67 &1.24 &1.38 \\
2024-03-23 &2460392.5  &  1.19 & oevc04030  & 0.56  &  1.75 & 1.32 & 1.25&1.47 &1.00 &1.26 \\
2024-09-27 &2460580.5  &  1.28 & oevc05030  & 0.59  &  1.66 & 1.30 & 1.18&1.46 &1.01 &1.31 \\
2024-11-17 &2460631.5  &  1.31 & ofai01030  & 0.58  &  1.63 & 1.17 & 1.04&1.33 &0.87 &1.10 \\
2024-12-20 &2460668.5  &  1.32 & ofai02030  & 0.59  &  1.49 & 1.08 & 1.08&1.30 &0.87 &1.06 \\
2025-08-27 &2460914.5  &  1.44 & ofgk01040  & 0.56  &  1.44 & 1.00 & 1.17&1.27 &0.91 &1.12 \\
2025-09-22&2460940.5   &  1.45 & ofgk02040  & 0.51  &  1.46 & 1.06 & 1.08&1.18 &0.84 &1.07 \\
2025-10-17&2460965.5   &  1.46 & ofgk03040  & 0.48  &  1.43 & 1.00 & 1.06&1.06 &0.81 &0.98 \\
2025-11-13&2460992.5   &  1.48 & ofgk04040  & 0.54  &  1.46 & 0.95 & 1.05&1.20 &0.83 &1.03    
\enddata

\end{deluxetable}


\begin{thebibliography}{}
\bibitem[Adams \& MacCormack(1935)]{1935ApJ....81..119A} Adams, W.~S. \& MacCormack, E.\ 1935, \apj, 81, 119. doi:10.1086/143620
\bibitem[Bernat \& Lambert(1975)]{1975ApJ...201L.153B} Bernat, A.~P. \& Lambert, D.~L.\ 1975, \apjl, 201, L153. doi:10.1086/181964
\bibitem[Chen et al.(2020)]{2020 ApJ...892..110C} Chen, Z., Ivanova, N.,\& Carroll-Nellenback, J.\ 2020, \apj, 892, 110. doi: 10.3847/1538-4357/ab7b6e
\bibitem[Cranmer \& Winebarger(2019)]{2019ARA&A..57..157C} Cranmer, S.~R. \& Winebarger, A.~R.\ 2019, \araa, 57, 157. doi:10.1146/annurev-astro-091918-104416
\bibitem[Dupree et al.(1987)]{1987ApJ...317L..85D} Dupree, A.~K., Baliunas, S.~L., Guinan, E.~F., et al.\ 1987, \apjl, 317, L85. doi:10.1086/184917
\bibitem[Dupree \& Stefanik (2013)]{2013EAS....60...77D}Dupree, A.~K. \& Stefanik, R. P.\ 2013, EAS Pub. Series, 60, 77. doi:10.1051/eas/1360008
\bibitem[Dupree et al.(2020)]{2020ApJ...899...68D} Dupree, A.~K., Strassmeier, K. G., Matthews, L. D., \etal\ 2020, \apj, 899, 68. doi:10.3847/1538-4357/aba516
\bibitem[Dupree et al.(2022)]{2022ApJ...936...18D} Dupree, A.~K., Strassmeier, K. G., Calderwood, T., \etal\ 2022, \apj, 936, 18. doi:10.3847/1538-4357/ac7853
\bibitem[Freytag \& H{\"o}fner(2008)]{2008A&A...483..571F} Freytag, B. \& H{\"o}fner, S.\  2008, \aap, 483, 571. doi:10.1051/0004-6361:20078096
\bibitem[Fuller \& Tsuna(2024)]{2024OJAp....7E..47F} Fuller, J. \& Tsuna, D.\ 2024, The Open Journal of Astrophysics, 7, 47. doi:10.33232/001c.120130
\bibitem[Gilliland \& Dupree (1986)]{1996ApJ...463L..29G} Gilliland, R. L. \& Dupree, A. K.\ 1996, \apjl, 463, L29. doi:10.1086/310043
\bibitem[Goldberg et al.(2024)]{2024ApJ...977...35G} Goldberg, J.~A., Joyce, M., \& Moln{\'a}r, L.\ 2024, \apj, 977, 35. doi:10.3847/1538-4357/ad87f4
\bibitem[Goldberg(1984)]{1984PASP...96..366G} Goldberg, L.\ 1984, \pasp, 96, 366. doi:10.1086/131347
\bibitem[Guinan(1984)]{1984LNP...193..336G} Guinan, E.~F.\ 1984, Cool Stars, Stellar Systems, and the Sun, 193, 336. doi:10.1007/3-540-12907-3\_225
\bibitem[Hagen et al.(1987)]{1987A&A...184..256H} Hagen, H.-J., Hempe, K., \& Reimers, D.\ 1987, \aap, 184, 256.   
\bibitem[Honeycutt et al.(1980)]{1980ApJ...239..565H} Honeycutt, R.~K., Kephart, J.~E., Bernat, A.~P., et al.\ 1980, \apj, 239, 565. doi:10.1086/158142 
\bibitem[Howell et al.(2025)]{2025ApJ...988...L47} Howell, S. B., Ciardi, D R., Clark, C. A., et al.\ 2025, \apjl, 988L, 47H.doi:10.3847/2041-821
\bibitem[Hummer \& Rybicki(1968)]{1968ApJ...153L.107H} Hummer, D.~G. \& Rybicki, G.~B.\ 1968, \apjl, 153, L107. doi:10.1086/180231
\bibitem[Joyce et al.(2020)]{2020ApJ...902...63J} Joyce, M., Leung, S.-C., Moln{\'a}r, L., et al.\ 2020, \apj,  902, 1, 63. doi:10.3847/1538-4357/abb8db
\bibitem[Kervella et al.(2018)]{2018A&A...609A..67K} Kervella, P., Decin, L., Richards, A. M. S. \etal\ 2018, \aap, 609, A67.  doi:10.1051/0004-6361/201731761
\bibitem[Kirsch \& Baade(1994)]{1994A&A...291..535K} Kirsch, T. \& Baade, R.\ 1994, \aap, 291, 535.  
\bibitem[Kiss et al.(2006)]{2006MNRAS.372.1721K} Kiss, L.~L., Szab{\'o}, G.~M., \& Bedding, T.~R.\ 2006,\mnras, 372, 1721. doi:10.1111/j.1365-2966.2006.10973.x
\bibitem[Kravchenko et al.(2021)]{2021A&A...650L..17K} Kravchenko, K., Jorissen, A., VanEck, S. \etal\ 2021, \aap, 650, L17.doi:10.1051/0004-6361/202039801
\bibitem[Lobel \& Dupree(2000)]{2000ApJ...545...454}Lobel, A. \& Dupree, A. K.\ 2000, \apj, 545, 454. doi: 10.1086/317784
\bibitem[Lobel \& Dupree(2001)]{2001ApJ...558...815} Lobel, A. \& Dupree, A. K.\ 2001, \apj, 558, 815. doi:10.1086/322284
\bibitem[Ma et al.(2025)] {2025arXiv251014875M } Ma, J-Z.,  Justham, S., Pakmor, R. et al.\ 2025, eprint arXiv:2510.14875. doi:10.48550/arXiv.2510.14875
\bibitem[MacLeod et al.(2025)]{2025ApJ...978...50M} MacLeod, M., Blunt, S., De Rosa, R.~J., et al.\ 2025, \apj, 978, 50. doi:10.3847/1538-4357/ad93c8
\bibitem[Montarg{\`e}s et al.(2014)]{2014A&A...572A..17M} Montarg{\`e}s, M., Kervella, P., Perrin, G., et al.\ 2014, \aap, 572, A17. doi:10.1051/0004-6361/201423538
\bibitem[Montarg{\`e}s et al.(2016)]{2016A&A...588A.130M} Montarg{\`e}s, M., Kervella, P., Perrin, G., et al.\ 2016, \aap, 588, A130. doi:10.1051/0004-6361/201527028
\bibitem[Montarg{\`e}s et al.(2021)]{2021Natur.594..365M} Montarg{\`e}s, M., Cannon, E., Lagadec, E., et al.\ 2021, \nat, 594, 7863, 365. doi:10.1038/s41586-021-03546-8
\bibitem[O'Gorman et al.(2020)]{2020A&A...638A..65O} O'Gorman, E., Harper, G. M., Ohnaka, K. et al.\  2020, \aap, 638,A65. doi:10.1051/0004-6361/202037756
\bibitem[Ostriker(1999)]{1999ApJ...513..252O} Ostriker, E.~C.\ 1999, \apj, 513, 1, 252. doi:10.1086/306858
\bibitem[Percy \& Sato(2009)]{2009JRASC.103...11P} Percy, J.~R. \& Sato, H.\ 2009, \jrasc, 103, 1, 11.
\bibitem[Raskin et al.(2011)]{2011A&A...526A..69R} Raskin, G., van Winckel, H., Hensberge, H., et al.\ 2011, \aap, 526, A69. doi:10.1051/0004-6361/201015435
\bibitem[Ridgway(2013)]{2013EAS....60....5R} Ridgway, S.~T.\ 2013, EAS Publications Series, 60, 5. doi:10.1051/eas/1360001
\bibitem[Soker(2021)]{2021ApJ...906....1S} Soker, N.\ 2021, \apj, 906, 1, 1. doi:10.3847/1538-4357/abca8
\bibitem[Soszy{\'n}ski et al.(2021)]{2021ApJ...911L..22S} Soszy{\'n}ski, I., Olechowska, A., Ratajczak, M., et al.\ 2021, \apjl, 911, L22. doi:10.3847/2041-8213/abf3c9
\bibitem[Stothers(2010)]{2010ApJ...725.1170S} Stothers, R.~B.\ 2010, \apj, 725, 1, 1170. doi:10.1088/0004-637X/725/1/1170
\bibitem[Stothers \& Leung(1971)]{1971A&A....10..290S} Stothers, R. \& Leung, K.~C.\ 1971, \aap, 10, 290
\bibitem[Uitenbroek et al.(1998)]{1998AJ....116.2501U} Uitenbroek, H., Dupree, A.~K., \& Gilliland, R.~L.\ 1998, \aj,  116, 5, 2501. doi:10.1086/300596
\bibitem[van der Hucht et al.(1980)]{1980A&A....82...14V} van der Hucht, K.~A., Bernat, A.~P., \& Kondo, Y.\ 1980, \aap, 82, 14.
\bibitem[Wedemeyer et al.(2017)] {2017A&A...606A..26W} Wedemeyer, S., Ku{\v{c}}inskas, A., Klevas, J., \& Ludwig, H-G.\ 2017, \aap, 606 A26: doi:10.1051/0004-6361/201730405
\bibitem[Weymann, R. (1962)]{1962ApJ...136..844W} Weymann, R.\ 1962, \apj, 136, 844. doi:10.1086/147441 
\bibitem[Wood et al.(2004)]{2004ApJ...604..800W} Wood, P.~R., Olivier, E.~A., \& Kawaler, S.~D.\ 2004, \apj, 604, 800. doi:10.1086/382123
\end{thebibliography}
\end{document}